\documentclass[twocolumn]{svjour3}          
\smartqed  
\usepackage{latexsym}
\usepackage{lipsum}
\usepackage{epsf}
\usepackage{float}
\usepackage{epsfig}
\usepackage{subfigure}
\usepackage{amsmath}
\usepackage{amssymb}
\usepackage{graphicx}
\usepackage{dcolumn}
\usepackage{bm}
\usepackage{xcolor}
\usepackage{tikz}
\usetikzlibrary{shapes}
\definecolor{blue}{RGB}{0,112,192}
\definecolor{lightblue}{RGB}{0,176,240}
\definecolor{green}{RGB}{0,176,80}
\definecolor{yellow}{RGB}{255,255,0}
\definecolor{orange}{RGB}{255,192,0}
\definecolor{red}{RGB}{255,0,0}
\definecolor{darkred}{RGB}{118,0,0}
\definecolor{purple}{RGB}{208,0,154}

\newcommand*{\tikzcirc}[2]{%
   \setbox0=\hbox{\strut}%
   \begin{tikzpicture}
     \useasboundingbox (-.25em,0) rectangle (.25em,\ht0);
     \filldraw[draw=#1,fill=#2] (0.0,0.35\ht0) circle[radius=.25em];
   \end{tikzpicture}%
}
\newcommand*{\tikzrect}[2]{%
   \setbox0=\hbox{\strut}%
   \begin{tikzpicture}
     \useasboundingbox (-.25em,0) rectangle (.25em,\ht0);
     \filldraw[draw=#1,fill=#2] (-0.25em,0.05em) rectangle (.25em,0.55em);
   \end{tikzpicture}%
}

\journalname{Granular Matter}
\begin{document}

\title{Deposition Morphology of Granular Column Collapses 
}


\author{Teng Man \and Herbert E. Huppert \and Ling Li \and Sergio A. Galindo-Torres}


\institute{ 
        T. Man \at Institute of Advanced Technology, Westlake Institute for Advanced Study, 18 Shilongshan St., Hangzhou, China \\
        School of Engineering, Westlake University, 18 Shilongshan St., Hangzhou, China \\
        \email{manteng@westlake.edu.cn}
        \and
        H. E. Huppert \at Institute of Theoretical Geophysics, King's College, University of Cambridge, King's Parade, Cambridge, United Kingdom
        \and
        L. Li \at School of Engineering, Westlake University, 18 Shilongshan St., Hangzhou, China
        \and
        S. A. Galindo-Torres \at School of Engineering, Westlake University, 18 Shilongshan St., Hangzhou, China\\
        \email{s.torres@westlake.edu.cn}
}

\date{Received: date / Accepted: date}

\maketitle

\begin{abstract}
Granular column collapses result in an array of flow phenomena and deposition morphologies, the understanding of which brings insights into studying granular flows in both natural and engineering systems. Guided by experiments, we carried out computational studies with the discrete element method (DEM) to identify fundamental links between the macro-scale behavior and micro-scale properties of granular columns. A dimensionless number combining particle and bulk properties of the column, $\alpha_{\textrm{eff}}$, was found from dimensional analysis to determine three collapse regimes (quasi-static, inertial, and liquid-like), revealing universal trends of flow regimes and deposition morphologies under different conditions. This dimensionless number may represent physically the competing inertial and frictional effects that govern the behavior of the granular column collapse, including energy conversion and dissipation. The finding is important for understanding quantitatively the flow of granular materials and their deposits.
\end{abstract}

\section{Introduction}
\label{intro}
Granular materials are used widely and encountered frequently in a variety of areas of civil engineering, geophysics, pharmaceutical engineering, food processing, and chemical engineering \cite{man2019rheology,Zhang2020coupling}. They can behave like a solid, a liquid, or a gas in different circumstances \cite{midi2004dense}. In recent decades, breakthroughs have been made to understand the basic governing principles, especially the constitutive relationships, of granular materials. Among these investigations, the proposal of the $\mu(I)$ rheology \cite{jop2006constitutive} and the $\mu(I, I_v)$ relationship \cite{trulsson2012transition,man2018rheology} of granular materials opened a window for exploring the behavior of dense granular materials from a viewpoint of a competition among acting stresses, viscous stresses and inertial stresses \cite{pouliquen2006flow,trulsson2012transition} (or a competition among different time scales $1/\dot{\gamma}$, $d/\sqrt{\sigma_n/\rho_p}$, and $\eta_f/\sigma_n$ \cite{cassar2005submarine}, where $\dot{\gamma}$ is the shear rate, $d$ the average particle radius, $\sigma_n$ the pressure, $\rho_p$ the particle density, and $\eta_f$ the dynamic viscosity of the interstitial fluid).

The collapse of granular columns has drawn continuous attention due to the potential link to the dynamics and deposition morphologies of various geophysical flows \cite{lube2005collapses,lagree2011granular}. Progress has been made, even though no universal relationship describing the collapse of granular columns has been concluded. Difficulties arise because such systems are highly dependent on inter-particle friction \cite{staron2007spreading} and boundary conditions \cite{staron2005study,lagree2011granular,zhang2014particle}, which were not fully captured by the rheological models previously mentioned. Some rigorous investigations \cite{lube2004axisymmetric,lajeunesse2005granular,lube2005collapses} have been conducted to understand the behavior of granular column collapses. Lube et al. \cite{lube2004axisymmetric,lube2005collapses} and Lajeunesse et al. \cite{lajeunesse2005granular} independently determined that relationships for both the normalized run-out distance $\mathcal{R}=(R_{\infty}-R_i)/R_i$ (where $R_{\infty}$ is the final radius of the granular pile, and $R_i$ the initial radius of the granular column), and the halt time of a collapsed granular column scale with the initial aspect ratio, $\alpha = H_i/R_i$ (where $H_i$ is the initial height) of the column, a parameter drawn from dimensional analysis. Lube et al. \cite{lube2004axisymmetric} further concluded that inter-particle friction only plays an important role in the last instant of the flow when the collapse starts to halt. 

Zenit \cite{zenit2005computer} simulated the collapse of 2-D granular columns and found that the 2-D collapse is essentially similar to that of axisymmetric granular column collapse performed by Lube et al. The dimensionless number, working in the axisymmetric granular column collapse, also works for the 2-D case. Staron and Hinch \cite{staron2007spreading} studied the effect of grain properties on the collapse of granular columns and suggested that the frictional coefficient could have a significant impact on the run-out distance and final mass distribution with 2-D granular column collapses, and found that, under certain circumstances, more particles may end up packing at the outer ring of the granular pile than packing in the center of it. Kermani et al. \cite{kermani2015simulation} used the discrete element method (DEM) to investigate the relationship between deposition morphology and inter-particle rotational resistance as well as the initial porosity of the granular packing, based on which they linked the deposition morphology to the energy dissipation. Lai et al. \cite{lai2017collapse} studied the collapse of granular column with various gain size distributions with experiments and simulations and show that particle flow mobility increases as the fractal dimension increases. The column collapse of more complex particles were also investigated rigorously that, e.g., Trepanier et al. \cite{trepanier2010column} studied the behavior of the granular column collapse with rod-like particles of different particle aspect ratios and showed that, when particle aspect ratio and the initial solid fraction reached certain levels, the granular packing become completely solid-like.

However, previous research often lacked physical interpretation of the scaling parameters, i.e. the initial aspect ratio $\alpha$ (which was derived from dimensional analysis \cite{lube2004axisymmetric}), and the universality of the transition point between the two determined relationships $\mathcal{R} \propto \alpha$ and $\mathcal{R} \propto \alpha^{1/2}$ \cite{lube2004axisymmetric,lube2005collapses}. Furthermore, previous studies did not quantitatively take the inter-granular friction and boundary conditions into consideration. In this paper, with the assistance of both experiments and DEM simulations, we aim to explore further the scaling of the final run-out distance of collapsed granular columns and link the scaling law with a theoretically derived dimensionless number. The deposition morphology and the correlation between initial height and deposition distance are carefully analyzed. Three different collapsing types are observed and further associated with the ratio between inertial stresses and frictional stresses during granular column collapses. With this work, the regime transition of the granular column collapse can be obtained for further investigation of geophysical flows, such as debris flows and landslides.

The paper is organized as follows. Section \ref{sec:simu} introduces the numerical method we use in this study and provides an experimental validation to show that the Voronoi-based discrete element method is able to capture the behavior of the collapse of dry granular columns. Section \ref{sec:res_n_disc} presents the simulation results and the corresponding analyses, within which we talked about the influence of inter-particle friction and particle/boundary friction on the run-out distance behavior as well as the deposition morphology of collapsed granular columns (Sect.\ \ref{sec:runout} and \ref{sec:morph}). We then introduced a physics-based dimensionless number to describe the behavior of granular column collapses. Further discussions are then presented in Sect. \ref{sec:dimensionAnal}, Sect. \ref{sec:alpha_eff} and Sect. \ref{sec:regime}, before conclusions are drawn in Sect.\ \ref{sec:conclu}.

\begin{figure}
    \centering
    \includegraphics[scale = 0.30]{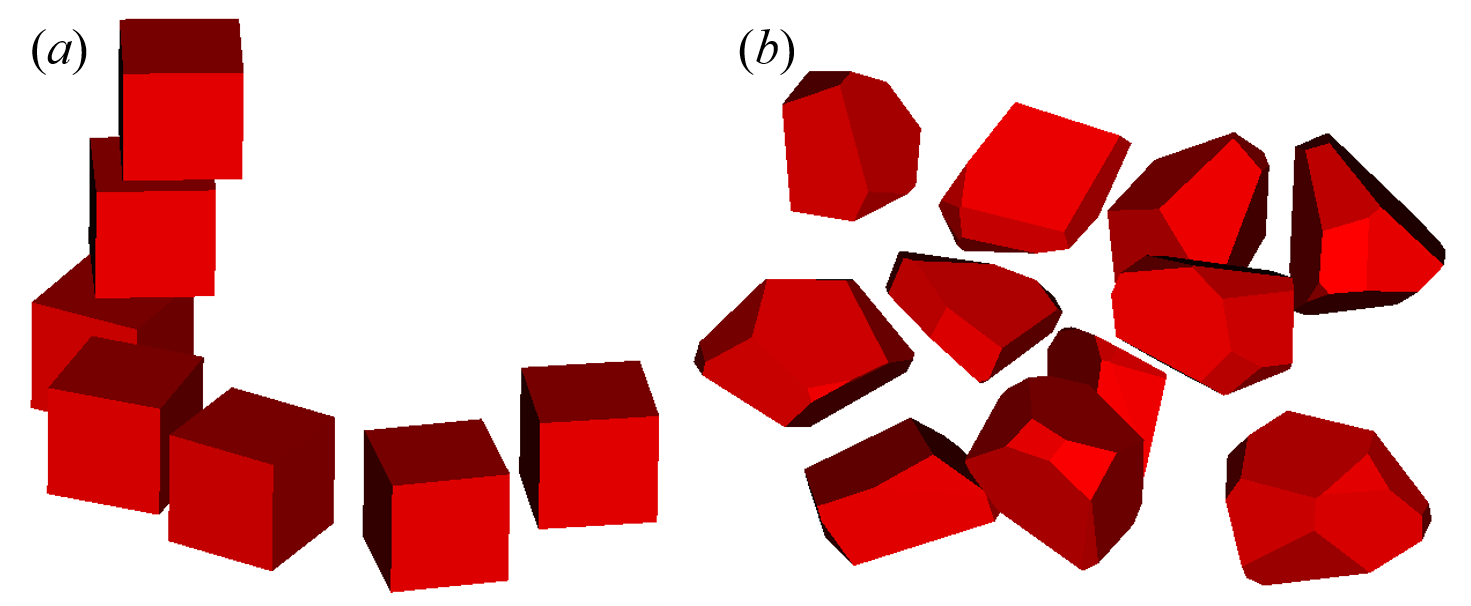}
    \caption{(a) and (b) show cubic particles and Voronoi-based particles, respectively.}
    \label{fig:particle}       
\end{figure}

\section{Simulations and experimental validation}
\label{sec:simu}

\begin{figure*}
\centering
\includegraphics[width=0.75\textwidth]{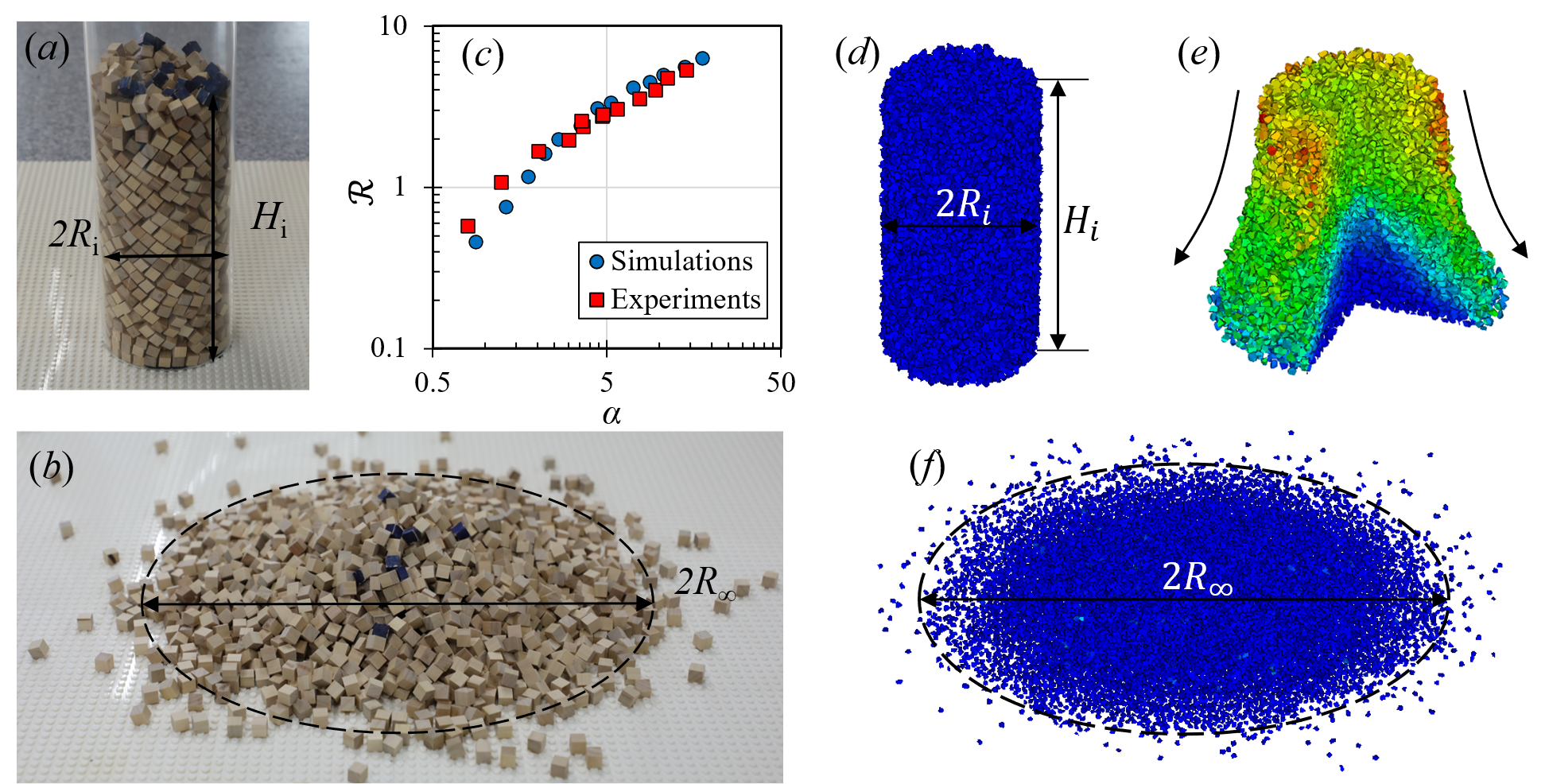}
\caption{(a) and (b) represent the experimental set-up and results of the collapse of dice; (c) shows the relationship between normalized run-out distance and the initial aspect ratio and the comparison between experimental results and the corresponding simulations; (d - f) The initial, intermediate, and final DEM simulations of granular column collapse. The color in Fig. (d-f) represents the velocity field, especially in Fig. (e), different colors (from blue to red) show a spectrum of velocity magnitude from 0 cm/s to approximately 10 cm/s}
\label{simu_setup}       
\end{figure*}

\subsection{Discrete element method}
\label{DEM}
In this study, we performed simulations with the discrete element method (DEM) \cite{cundall1979discrete,galindo2009molecular,galindo2010molecular} to test the collapse of granular columns, which allowed us to extract particle-scale data from the system and calculate the dynamics of particles with Newton's laws which are integrated numerically for both translational and rotational degrees of freedom. We use sphero-polyhedron particles in our simulations. The sphero-polyhedra method was initially introduced by Pourning\cite{pournin2005generalization} for the simulation of complex-shaped DEM particles. Later, it was modified by Alonso Marroquin\cite{alonso2008spheropolygons}, who introduced a multi-contact approach in 2D allowing the modelling of non-convex shapes and was extend to 3D by Galindo-Torres et. al.\cite{galindo2009molecular}. A sphero-polyhedron is a polyhedron that has been eroded and then dilated by a sphere element. The result is a polyhedron of similar dimensions but with rounded corners.

The best advantage of the sphero-polyhedra technique is that it allows for an easy and efficient definition of contact laws between the particles. This is due to the smoothing of the edges of all geometric features by circles (in 2D) or spheres (in 3D). Regarding the contact between two generic particles $P_1$ and $P_2$, first one has to consider the contact between each geometric feature of particle $P_1$ with all features of particle $P_2$. In mathematical notation, both $P_1$ and $P_2$ have a set of vertices $\{V_{1,2}^i\}$, edges $\{E_{1,2}^j\}$, and faces $\{F_{1,2}^k\}$. Thus, a particle is defined as a polyhedron, i.e. a set of vertices, edges and faces, where each one of these geometrical feature is dilated by a sphere.

For simplicity, We denominate the set of all the geometric features of a particle as
$\{G_{1,2}^i\}$. Now the function representing this topology can be defined as the
distance function for two geometric features according to
\begin{equation}\label{eq:dist}
 \textrm{dist}(G_1^i,G_2^j) = \min\left (\textrm{dist}(\vec{X_i},\vec{X_j})\right ),
\end{equation}
where $\vec{X_i}$ is a 3D vector that belongs to the set $G_1^i$. This means that the
distance for two geometric features is the minimum Euclidean distance assigned to two
points belonging to them.

\begin{figure}[H]
    \centering
    \includegraphics[scale = 0.5]{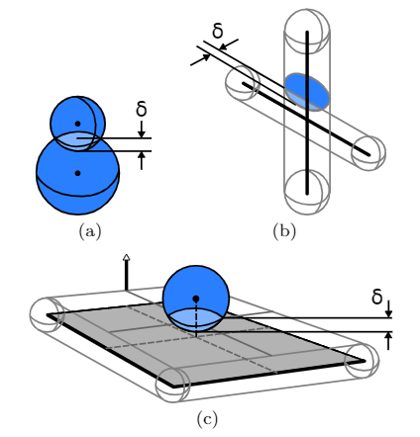}
    \label{fig:deltas}
\end{figure}

Since both particles are dilated by their sphero-radii $R_1$ and $R_2$, it can be said
that there is a complete contact when the distance between the two geometric features is less than the addition of the corresponding radii used in the sweeping stage, i.e.:
\begin{equation}\label{eq:contact}
 \textrm{dist}(G_1^i,G_2^j) < R_1+R_2,
\end{equation}
and the corresponding contact overlap $\delta_n$ can be calculated accordingly. Here, the advantage of the sphero-polyhedra technique becomes evident since this
definition is similar to the one for the contact law of two spheres\cite{belheine2009numerical}. We note that, in our simulation, three types of contacts (vertex-vertex contact, edge-edge contact, and vertex-face contact, shown in Fig. \ref{fig:deltas}) are considered. For these types of contacts, we implement a Hookean contact model with energy dissipation to calculate the interactions between particles. At each time step, the overlap between adjacent particles, $\delta_n$, is checked and the normal contact force can be calculated using
\begin{equation}
    \begin{split}
        \vec{F}_n = -K_n\delta_n\hat{n} - m_e\gamma_{n}\vec{v}_n,
    \end{split}
\end{equation}
where $K_n$ is the normal stiffness characterizing the deformation of the material, $\hat{n}$ is defined as the normal unit vector at the plane of contact, $\vec{v}_n$ is the relative normal velocity between particles, $m_e = 0.5(1/m_1 + 1/m_2)$ is the reduced mass of the contacting particle pair, and $m_1$ and $m_2$ are masses of contacting particles, respectively, and $\gamma_n$ is the normal energy dissipation constant, which depends on the coefficient of restitution $e$ as \cite{alonso2013experimental,galindo2018micromechanics},
\begin{equation}
    \begin{split}
        e = \textrm{exp}\left(-\frac{\gamma_n}{2} \frac{\pi}{\sqrt{\frac{K_n}{m_e} - (\frac{\gamma_n}{2})^2}}\right)\ .
    \end{split}
\end{equation}

The tangential contact forces between contacting particles were calculated by keeping track of the tangential relative displacement $\vec{\xi} = \int \vec{v}_tdt$. Thus, the tangential contact forces follow
\begin{equation}
    \begin{split}
        \vec{F}_t = -\textrm{min}\left(|K_t\vec{\xi}|,\ \mu_p |\vec{F}_n|\right)\hat{t},
    \end{split}
\end{equation}
where $K_t$ is the tangential stiffness, $\hat{t}$ is the tangential vector in the contact plane and parallel to the tangential relative velocity, $\vec{v}_t$, and $\mu_p$ is the frictional coefficient between contacting particles and can be replaced by the frictional coefficient between the particles and the bottom boundary, $\mu_w$, while calculating the particle-boundary interactions. In this study, since we use either cubic particles or Voronoi-based particles, no rolling resistance is needed to be considered. The motion of particles is then calculated by step-wise resolution of Newton's second law with the normal and contact forces mentioned before. The same neighbor detection and force calculation algorithms were already been discussed and validated in previous studies. The presented DEM formulation has been validated before with experimental data \cite{cabrejos2016assessment,belheine2009numerical} and is included in the MechSys open source multi-physics simulation library \cite{galindo2013coupled}. In this study, we use two kinds of particles: (1) cubic particles [Fig. \ref{fig:particle}(a)] and (2) Voronoi-based sphero-poly-hedron particles [Fig. \ref{fig:particle}(b)]. The complex-shaped particles are modeled with the concept of sphero-polyhedron, which is the object resulting from dilating a polyhedron by a sphere. The overlap between contacting particles and contact force resulting from it can be then calculated \cite{galindo2009molecular,galindo2010molecular}.

\subsection{Experimental validations}

Before diving deeply into the simulation setup, we first validated our DEM model with experiments. We performed experiments on the collapse of granular columns with wooden dice [Fig. \ref{simu_setup}(a) and (b)] of length 1.0 cm. The density of the dice was 0.681 g/cm$^3$. They were dropped randomly into a circular tube ($D = 2R_i = 11.4$ cm, where $D$ is the internal diameter of the cross-section of the plastic tube, and $R_i$ is the internal radius of the tube.). In the experiments, we used a LEGO$\circledR$ board [Fig. \ref{simu_setup}(a)] as the bottom plate. The frictional coefficient between the board and the wooden dice was measured to be approximately 0.4, and the frictional coefficient between the wooden dice 0.84.

After placing the dice into the plastic tube [Fig. \ref{simu_setup}(a)], we measured the initial height of the granular packing, $H_i$. Then, the tube was manually lifted to release all the particles to form a granular pile. The run-out distance was measured as shown in Fig. \ref{simu_setup}(b). Then, we could measure the final deposition radius, $R_{\infty}$, of the granular pile and calculate the normalized run-out distance and the corresponding initial aspect ratio with the following equation,
\begin{subequations} \label{eq:normalized_runout}
\begin{align}
    \mathcal{R} &= \frac{R_{\infty} - R_i}{R_i}\ ,\\
    \alpha &= \frac{H_i}{R_i}\ .
\end{align}
\end{subequations}
Thus, as we vary the initial height, $H_i$, of the granular column, the relationship between $\mathcal{R}$ and $\alpha$ was obtained [\ \tikzrect{black}{red} in Fig. \ref{simu_setup}(c)]. Because the tube was lifted manually, a disturbance would be applied to the granular pile while lifting, thus giving a relatively higher run-out distance when the initial aspect ratio is small.

We then performed numerical simulations with the same set-up as in the experiments of wooden dice [the DEM elements are cubic particles, as shown in Fig. \ref{fig:particle}(a)], to make sure our simulation set-up could capture the behavior of granular column collapses well. We varied the initial height of the granular column between 5 cm and 100 cm while keeping the initial cross-section diameter constant at 11.4 cm. In both experiments and the corresponding simulations, the number of particles varied from approximately 200 (the shortest column) to approximately 4500. In this set of simulations, we set the inter-particle frictional coefficient $\mu_p = 0.84$, particle/boundary frictional coefficient $\mu_w = 0.4$, and restitution coefficient $e = 0.2$. Blue markers in Fig. \ref{simu_setup}(c) show the results of this set of simulations. This shows that our model with a simple contact law represents the experimental data well, especially when the initial aspect ratio is large. When the initial aspect ratio is small, the experimental results tend to be larger than the simulation results. This might be due to the disturbance generated while manually lifting the tube during experiments. when the aspect ratio is small, the effect of lifting the tube during the experiments had significant influence on the run-out distance. It could also be that the frictional coefficient of dice/board contact is more complex than being generalized as one frictional coefficient, and we  may have overestimated the frictional coefficient when the aspect ratio is small, which results in smaller run-out distance.

\subsection{Simulation setup}

Then, to investigate the behavior of granular column collapses and the resulting run-out distances and deposition morphologies, we performed simulations of the granular column collapses with Voronoi-based sphero-polyhedra\cite{galindo2009molecular,galindo2010molecular}. We note that the shape of particles (Voronoi-based particles, dice, ellipsoids, spheres, particles with different thickness/length aspect ratios, etc.) could significantly influence of deposition morphology, in this study, we focus on using Voronoi-based particles to investigate a general behavior similar to that of sand particles. The detailed influence of particle shapes on the granular column collapses will be further explored in the future. The number of particles within unit length (1.0 cm) is 5, so the average particle size is $\approx 2$ mm. Particles were packed within a column of radius $R_i$ equal to 2.5 cm and varying heights $H_i$ leading to cases of different initial aspect ratio $\alpha$ [Fig. \ref{simu_setup}(d)]. Then, 20\% of the sphero-polyhedron particles were removed to form a packing with a solid fraction of $\phi_s =$ 0.8. $H_i$ varies from 1 cm to 50 cm. In the simulations of Voronoi-based particles, the number of particles varied from approximately 1900 to approximately 98500. Then, we removed the cylindrical tube in the simulation and let grains flow downward freely with the gravitational acceleration. Figure \ref{simu_setup}(e) shows the behavior of a granular column in the middle of the collapse. We can see that, the inner part of the column remained stationary, while the outer part of the column started to flow downward and spread outward. In the end, a cone-like pile of granular material with packing height, $H_{\infty}$, and average packing radius, $R_{\infty}$, will form [Fig. \ref{simu_setup}(f)]. 

We implemented the Hookean contact model (elaborated in Sect. \ref{DEM}) with energy dissipation and restitution coefficient $e = 0.1$ to calculate the interactions between particles as we have described in the above section. A relatively low value of $e$ was chosen to represent the rough surface of particles in real conditions\cite{li2020surface}. Simulations were conducted with varied initial aspect ratios, $\alpha$, between 0.4 and 20, varied inter-particle frictional coefficients, $\mu_p =$ 0.1, 0.2, 0.3, 0.4, 0.6, 0.8, and particle/boundary frictional coefficients, $\mu_w =$ 0.2, 0.3, 0.4, 0.6, 0.8. Based on these simulations we obtained the run-out behavior and deposition morphology for different conditions.

\subsection{Measurement of the run-out distance}

\begin{figure}
  \centering
  \includegraphics[scale = 0.35]{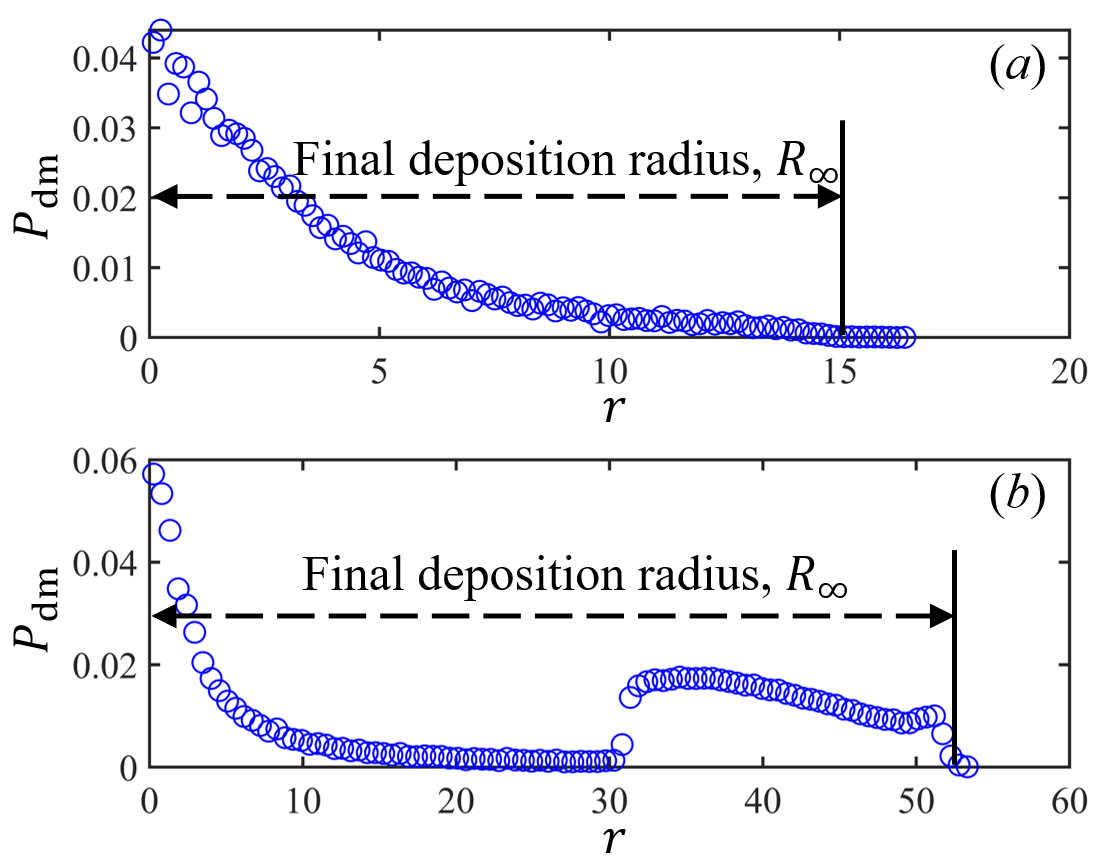}
  \caption{Method for measuring the run-out distance of a collapsed granular column. Here we use the radial histogram of particles to determine the run-out distance of a collapsed granular column. The $x-$axis is the radial position $r$, and the $y$-axis is the percentage of number of particles located within $(r-\Delta r/2, r+\Delta r/2)$ divided by the radial position. (a) $H_i = 4$ cm, (b) $H_i = 40$ cm}
  \label{runout_measure}
\end{figure}

In experiments, since the particles used were wooden dice, the collapsed granular column could form a well-packed granular pile with an almost-circular boundary. We measured the diameters of the final pile along 4 different directions and took the average and divided it by 2 to obtain the final deposition radius. In simulations, the measurement of the final radius is more complicated than that in experiments. In cases with small particle/boundary and inter-particle frictional coefficients but large initial aspect ratios, the spread of particles is far-reaching and leads to sparse (single layer) coverage of the area, especially at the front edge. In these cases, it is difficult to determine the edge/boundary and hence the final run-out distance. Thus, we measured the final radius with a histogram of particle distribution for each simulation. The following figure (Fig. \ref{runout_measure}) gives us two examples of how we measured the run-out distance.

In this figure, the x-axis is the radial position $r$, and the $y$-axis is the percentage of number of particles located within $(r-\Delta r/2, r+\Delta r/2)$ divided by the radial position, 
\begin{equation} \label{eq:runout_measure}
    \begin{split}
        P_{\textrm{dm}}(r) = \frac{1}{r}\left[N\left(r - \frac{\Delta r}{2}, r + \frac{\Delta r}{2}\right)/\left(\Sigma_rN\right)\right]\ ,
    \end{split}
\end{equation}
where $\Delta r$ is the bin width of the histogram, and $N(r-\Delta r/2, r+\Delta r/2)$ is the number of particles located between $r-\Delta r/2$ and $r+\Delta r/2$, and $\Sigma_rN$ is the total number of particles in one simulation. Fig. \ref{runout_measure}(a) shows the normalized particle number distribution (i.e. deposition morphology) of a simulation with $\mu_w=0.2$, $\mu_p=0.1$, and $H_i=4$ cm. It shows that most particles locate within $r \leq 15$ cm; thus, we take $R_{\infty} = 15$ cm. Fig. \ref{runout_measure}(b) displays the particle distribution of a simulation with $\mu_w=0.2$, $mu_p=0.1$, and $H_i=50$ cm.  The results show that, besides particles located in the middle of the final granular pile, a large number of particles are located at the front, and when $r>52$ cm, $P_{\textrm{dm}}(r)$ became 0. Thus, the final radius of the collapsed granular column is 52 cm in this example.

\section{Results and discussions}
\label{sec:res_n_disc}

\subsection{Run-out distance of granular column collapses}
\label{sec:runout}

\begin{figure}
  \centering
  \includegraphics[scale = 0.35]{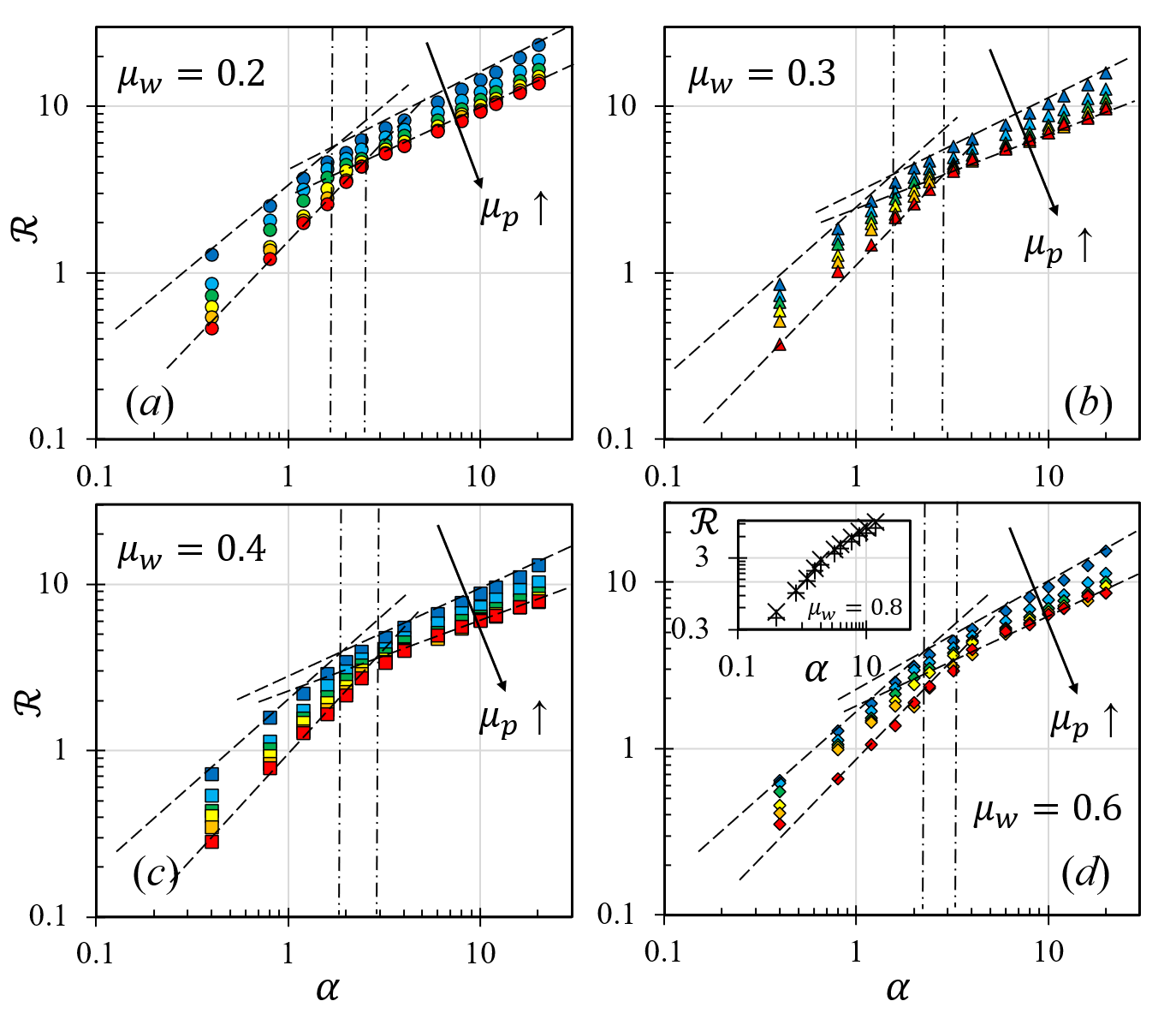}
  \caption{(a) - (d) and the insert of Fig. (d) show the relationship between the normalized run-out distance, $\mathcal{R} = (R_{\infty} - R_i)/R_i$, and the initial aspect ratio, $\alpha = H_i/R_i$, with the particle/boundary frictional coefficient equal to 0.2 (circle markers), 0.3 (triangular markers), 0.4 (square markers), 0.6 (diamonds), 0.8 [$\times$ for $\mu_p = 0.2$ and $+$ for $\mu_p = 0.4$ on the insert of Fig. (d)], respectively. Different color represents different inter-particle frictional coefficients. In Fig. (a), \protect\tikzcirc{black}{blue}: $\mu_p = 0.1$, \protect\tikzcirc{black}{lightblue}: $\mu_p = 0.2$, \protect\tikzcirc{black}{green}: $\mu_p = 0.3$, \protect\tikzcirc{black}{yellow}: $\mu_p = 0.4$, \protect\tikzcirc{black}{orange}: $\mu_p = 0.6$, \protect\tikzcirc{black}{red}: $\mu_p = 0.8$. The dashed lines indicate the fitting power-law relations of cases with most frictional particles and least frictional particles in each sub-figure. The dash-dot lines show the transition points of cases with most frictional particles and least frictional particles in each sub-figure}
  \label{results}
\end{figure}

Figure \ref{results} shows relationships between normalized run-out distance $\mathcal{R}$ and the initial aspect ratio $\alpha$. It shows that both the inter-particle friction and the particle-boundary friction play important roles in the run-out behavior. Similar to results obtained by other research, when $\alpha$ is sufficiently small, $\mathcal{R}$ scales roughly proportional to $\alpha$ ($\mathcal{R} \propto \alpha$). When $\alpha$ is larger than a threshold, $\mathcal{R}$ scales approximately proportional to $\alpha^{1/2}$ ($\mathcal{R} \propto \alpha^{1/2}$).

In each of these figures, as we increase the inter-particle friction, the run-out distance decreases since the additional friction increases the energy dissipation during the collapse. Similar results can be obtained when we increase particle/boundary friction. Although qualitative observations have been witnessed by others, the quantitative influence of inter-particle and particle/boundary friction on the flowing behavior of collapsed granular columns has been absent from previous studies. Also, as we vary the frictional coefficient, the point where the slope changes in the log-log plot of the $\mathcal{R}-\alpha$ relationship shifts (dash-dot lines in Fig. \ref{results}), which indicates that the change of collapse regimes does not only depend on the $\alpha$ of the granular column.

\subsection{Deposition morphology}

\label{sec:morph}
\begin{figure*}
  \centering
  \includegraphics[width=0.75\textwidth]{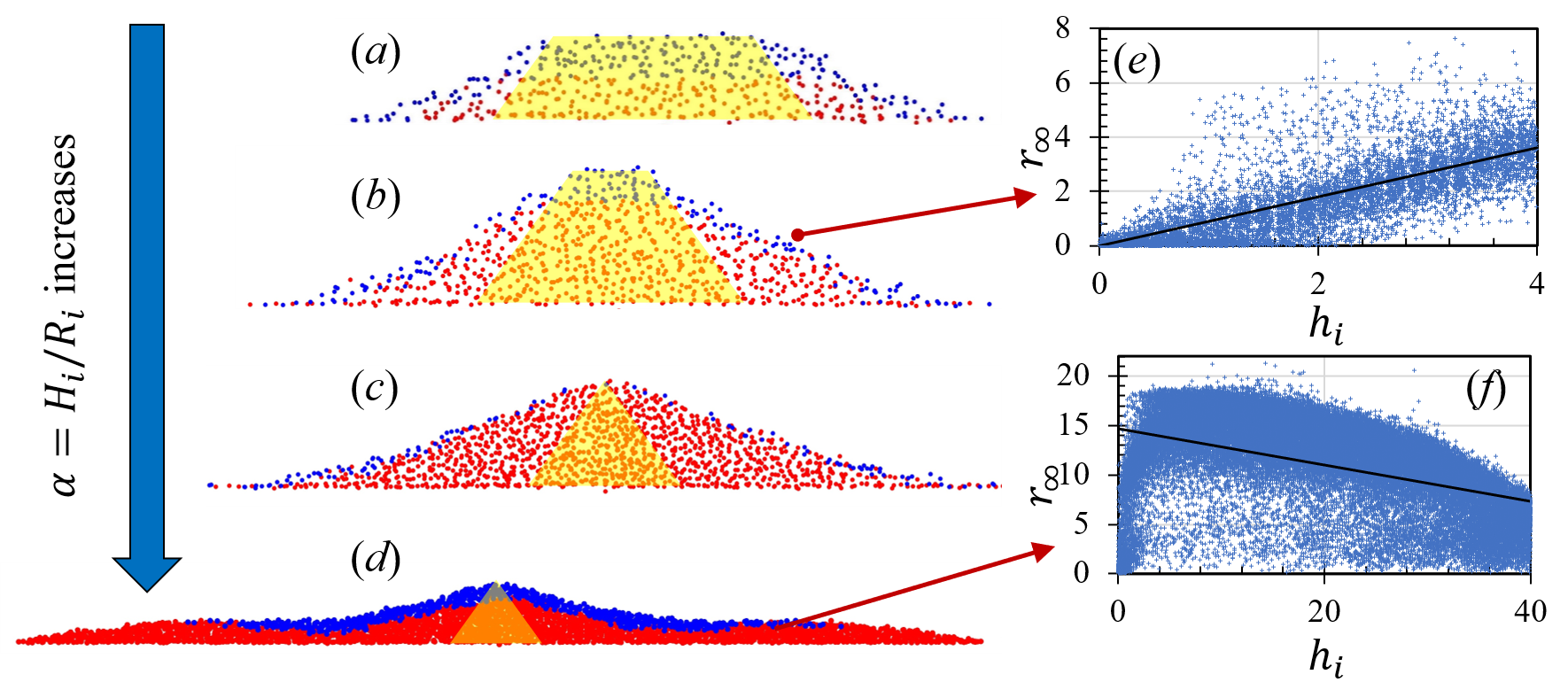}
  \caption{(a) - (d) Deposition morphology of collapsed granular columns. Blue dots denote particles originally at the top 25\% of the column, while red dots represent the rest. The yellow shaded areas roughly represent particles that remain stationary during collapses. (a) - (d) correspond to collapses of granular columns with $\alpha = 0.4, 1.2, 3.2, 10$, respectively. (e) and (f) show the relationship between the initial height $h_i$ and the final radial deposition $r_{\infty}$ of each particle. The unit of the axes in (e) and (f) is centimeters}
  \label{deposition}
\end{figure*}

To reveal the deposition morphology, we plot in Fig.\ \ref{deposition}(a)-(d) the deposition shapes of collapsed granular col-umns with different initial aspect ratios. We also distinguish particles initially located on the top of the column (blue dots) from the rest (red dots), while shading relatively stationary areas in yellow. The results show three different types of deposition morphology. When $\alpha$ is small [Fig.\ \ref{deposition}(a) and (b)], there is a plateau on top of the granular pile, and all blue dots present on top of red dots. In these cases, a large portion of particles remain stationary (yellow shaded area) during the collapse process, and the granular collapse behaves more like a yielded solid. Thus, we call it the quasi-static collapse. As we increase the initial aspect ratio [Fig.\ \ref{deposition}(c)], more particles flow down and spread out from the top of the column, and the inertial effect of particles starts to overtake the resistance of friction and become the dominating contribution. In these cases, top particles tend to rest at the foot of the final granular pile and during the collapse, surface granular flows (granular avalanches) dominate the behavior of the granular collapse. The final deposition morphology is similar to a typical sand pile. Thus, we name this type the inertial collapse, since inertial effects are dominating the macroscopic behavior. As we further increase $\alpha$, the deposition surface changes to Fig. \ref{deposition}(d). Particles initially on the top end up being in the center of the final granular pile, indicating that during the granular collapse, the top particles are flowing downward while pushing the lower particles out, and the flow behaves more like a liquid. We call this type of granular collapse a liquid-like collapse.

We can also see the transition in the plot of the relationship between the initial height $h_i$ and the final radial position $r_{\infty}$ of each particle [Fig. \ref{deposition}(e) and (f)]. When $\alpha$ is small [Fig. \ref{deposition}(e)], particles initially on the top tend to collapse to the front. Thus, the two parameters have a positive correlation. When $\alpha$ is large [Fig. \ref{deposition}(f)], particles initially on the top no longer reach the front of the final deposition. Thus, the two parameters have a negative correlation.

As we have described, these three morphologies correspond to three collapsing regimes: 
\begin{itemize}
    \item \textbf{a quasi-static regime}, where most bulk materials remain relatively stationary, and a plateau can form after the collapse;
    \item \textbf{an inertial regime}, where particles initially at the top end up flowing to the front of the final pile;
    \item \textbf{a liquid-like regime}, where the collapsed granular materials flow like a liquid to form a much more complex morphology, where particles initially at lower height start to accumulate at the front of the flow due to the push-out effect from particles at higher initial heights. 
\end{itemize}

We will elaborate the classification of three different collapse regimes of granular column collapses in following sections.

\subsection{Dimensional analysis}
\label{sec:dimensionAnal}

\begin{figure*}
  \centering
  \includegraphics[width=0.75\textwidth]{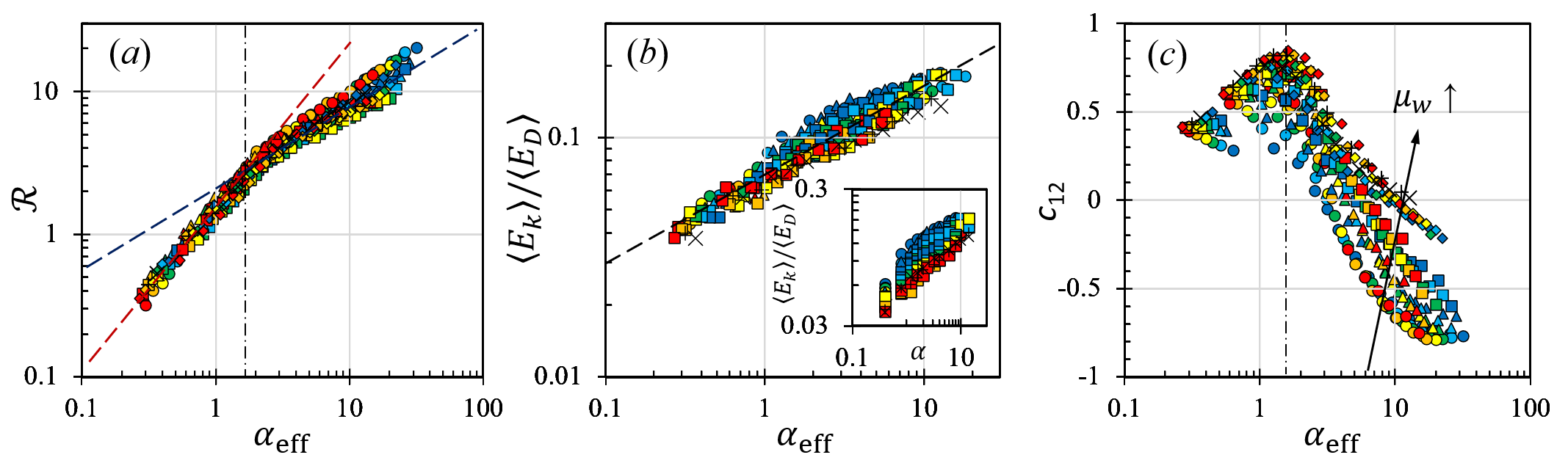}
  \caption{(a) shows the relationship between $\mathcal{R}$ and $\alpha_{\textrm{eff}} = (\sqrt{1/(\mu_w + \beta\mu_p)})(H_i/R_i)$. (b) presents $\langle E_k\rangle/\langle E_D\rangle$ against $\alpha_{\textrm{eff}}$, where $\langle E_k\rangle$ is the time averaged kinetic energy generated during the granular collapse, and $\langle E_D\rangle$ is the time averaged dissipated energy. The inset of (b) plots $\langle E_k\rangle$ against $\alpha$ in (c), the $y-$axis represents correlation coefficients between the initial height of particles and run-out distances of them, $c_{12} = \textrm{cov}[h_i, ({r}_{\infty} - {r}_i)]/(\sigma_h \sigma_{r})$, where $\textrm{cov}[]$ is the covariance function, and the $x-$axis shows $\alpha_{\textrm{eff}}$. The markers have the same meaning as that in Fig. \ref{results}. The dashed lines in Fig. (a) and (b) are fitted power-law relations and the dash-dot lines in Fig. (a) and (c) indicate regime changes}
  \label{effectRatio}
\end{figure*}

We consider the governing equation of the dynamics of a single particle in a granular system to be
\begin{equation}
    \begin{split}
        m_p\frac{d}{dt}\left(\frac{d\vec{x}}{dt}\right)=F_n\hat{n} + F_t\hat{t} + m_pg\hat{z}\ ,
    \end{split}
\end{equation}
where $m_p$ is the particle mass, $F_n$ and $F_t$ are normal and tangential contact forces acting on a particle, $\hat{z}$ is the unit vector in the direction of gravitational acceleration. We non-dimensionalize the governing equation with particle size $d_p$, gravitational acceleration $g$, column height $H_i$, and column radius $R_i$ so that,
\begin{equation}
    \begin{split}
        t^* = \frac{t\sqrt{gH_i}}{H_i},\ \vec{v}^* = \frac{\vec{v}}{\sqrt{gH_i}},\ \vec{x}^* = \frac{\vec{x}}{R_i},
    \end{split}
\end{equation}
where $t^*$, $\vec{v}^*$, and $\vec{x}^*$ are dimensionless time, dimensionless velocity, and dimensionless position vector, respectively. The dimensionless mass is $m^* = m_p/(\rho_pd_p^3)$. The normal force acting on particles can be non-dimensionalized using,
\begin{equation}
    \begin{split}
        F_n^* = \frac{F_n}{f(\phi_s) H_i R_i^2\rho_p g},
    \end{split}
\end{equation}
where $f(\phi_s)$ is a function of the solid fraction and we simplify the tangential force acting on particles as ${F}_t = f_1(\mu)|{F}_n|$, thus, The combination of normal and tangential forces can be written in the following form,
\begin{equation}
    \begin{split}
        F_n\hat{n} + F_t\hat{t} = f(\phi_s)g(\mu)\rho_p gH_iR_i^2 F_n^*\hat{s},
    \end{split}
\end{equation}
where $g(\mu) = \sqrt{1+[f_1(\mu)]^2}$ is a function of the frictional coefficient, $\hat{s}$ is the unit vector along the summation vector of the normal and tangential forces, and the governing equation can be non-dimensionalized in the following way,
\begin{equation}
    \begin{aligned}
    \rho_{p}d_p^3g\frac{R_i}{H_i}m^*\frac{d^2\vec{x}^*}{d{t^*}^2} &= f(\phi_s)g(\mu)\rho_p gH_iR_i^2 F_n^*\hat{s} \\
     &\ \ \ + \rho_pd_p^3g m^*\hat{z}\ .
    \end{aligned}
\end{equation}

We divide the governing equation by the parameter of the left-hand side of the equation and could obtain two dimensionless numbers in front of the two term on the right-hand side of the equation,
\begin{subequations}
\begin{align}
    \mathcal{I}_1 &= f(\phi_s)g(\mu)\left( \frac{H_i}{R_i}\right)^2 \left( \frac{R_i}{d_p}\right)^3,\\
    \mathcal{I}_2 &= \frac{H_i}{R_i},
\end{align}
\end{subequations}
where $\mathcal{I}_2$ coincides with the initial aspect ratio $\alpha$ and $\mathcal{I}_1$ combines the influence of the initial solid fraction, the friction properties, and the size effect $R_i/d_p$. This implies that, in addition to $\alpha$, we may come up with another dimensionless number to describe the behavior of granular column collapses. It is noted that we do not consider the size effect in this study and $R_i/d_p$ remains constant. The detailed analysis of $\mathcal{I}_1$ and the relatively optimal form of it is discussed in the following section.

\subsection{Effective aspect ratio}
\label{sec:alpha_eff}

In previous section, we find that the granular column collapse might be governed by both $\mathcal{I}_1$ and $\mathcal{I}_2$, while previous research argued that the run-out distance of a collapsed dry granular column is controlled almost entirely by the initial aspect ratio with little qualitative influence of friction. The square root of $\mathcal{I}_1$ can be seen as an effective aspect ratio. Since in this study we do not consider the size effect and the influence of the initial solid fraction, the effective aspect ratio, $\alpha_{\textrm{eff}}$, can be reduced to the following form,
\begin{equation} \label{eq:effect_alpha}
    \alpha_{\textrm{eff}} = \sqrt{g(\mu)}\left(\frac{H_i}{R_i}\right)\ \ .
\end{equation}
where $\mu$ is a general frictional coefficient of the granular system, which can be considered as a combination of both the inter-particle friction and the particle-boundary friction. We find that, to collapse all the data in the relationship between the normalized run-out distance and the effective aspect ratio, $g(\mu)$ should take the form of
\begin{equation} \label{eq:g_mu}
    \begin{split}
        g(\mu) = \frac{1}{\mu_w + \beta\mu_p}\ ,
    \end{split}
\end{equation}
where $\mu_w$ is the frictional coefficient between particles and the bottom boundary, $\mu_p$ is the frictional coefficient between particles, and $\beta$ is a fitting parameter, and $\beta = 2.0$ best fits the simulation data in Fig. \ref{effectRatio}(a) where we plot the relationship between the normalized run-out distance $\mathcal{R}$ and the effective aspect ratio $\alpha_{\rm{eff}}$ [thus, $\alpha_{\textrm{eff}}=(1/\sqrt{\mu_w+2\mu_p})(H_i/R_i)$ and the collapsed $\mathcal{R}(\alpha_{\rm{eff}})$ relationship is not sensitive to $\beta \in (1.5, 5.0)$]. The parameter $\beta$ can be roughly seen as a ratio between energy dissipated by inter-particle friction to that by particle/boundary friction. $\beta = 2.0$ indicates that the particle/boundary friction-induced energy dissipation plays a lesser role in the frictional energy dissipation. With $\beta = 2.0$, simulation results with different inter-particle friction and particle/boundary friction almost all collapse onto a single master curve, except when $\mu_w = 0.2$ and $\alpha_{\textrm{eff}} \gtrapprox 3.0$. In contrast to previous studies, for which $\alpha$ is the only criterion to indicate a regime change [where the slope of $\mathcal{R}(\alpha)$ changes] and no influence of frictional coefficient was quantitatively considered, we find that, as depicted in Fig. \ref{effectRatio}(a), the slope change in the $\mathcal{R}(\alpha)$ relationship occurs at a unique value of $\alpha_{\rm{eff}}$ for all cases. This indicates that $\mathcal{R}$ is almost entirely controlled by $\alpha_{\rm{eff}}$, which includes the influence of both inter-particle and particle/boundary friction, rather than $\alpha$ alone. The reason why run-out distances for $\mu_w = 0.2$ deviate from other cases is that under low particle-boundary friction, granular columns tend to form only one layer of particles with few inter-particle contacts, thus increasing the scattering condition. In other cases, particles have a more resistant reaction with the boundary condition and adjacent particles, where the kinetic energy of particles can be quickly dissipated. 

We also analyze the data from the viewpoint of energy consumption. For each simulation, we obtained the kinetic energy at time $t$, $E_k(t)$ and calculated the time averaged kinetic energy using
\begin{equation}
    \begin{split}
        \langle E_k\rangle = \frac{1}{\tau}\int_0^{\tau}E_k(t)dt\ ,
    \end{split}
\end{equation}
where $\tau$ is the terminating time of the granular column collapse. The kinetic energy can be seen as the trace of the inertial stress generated during the collapse of granular column. We then calculated the time averaged dissipated energy 
\begin{equation}
    \begin{split}
        \langle E_D\rangle = \frac{1}{\tau}\int_0^{\tau}\left[E_p(t)+E_k(t)-E_p(0)\right]dt\ ,
    \end{split}
\end{equation}
where $E_{p}(t)$ is the potential energy of the system at time $t$ and $E_{p}(0)$ is the initial potential energy. We plot the ratio between $\langle E_k\rangle$ and $\langle E_D\rangle$ with respect to $\alpha_{\textrm{eff}}$ in Fig.\ \ref{effectRatio}(b), which shows that $\langle E_k\rangle/\langle E_D\rangle$ has a power-law correlation with $\alpha_{\textrm{eff}}$, whereas the data points fail to collapse onto a power-law relation when we use $\alpha$ as the $x$-axis [inset of Fig. \ref{effectRatio}(b)]. This indicates that the effective aspect ratio is linked not only to the ratio between inertial stresses and frictional stresses, but also to the energy consumption ratio during the collapse.

\begin{figure}
  \centering
  \includegraphics[scale = 0.32]{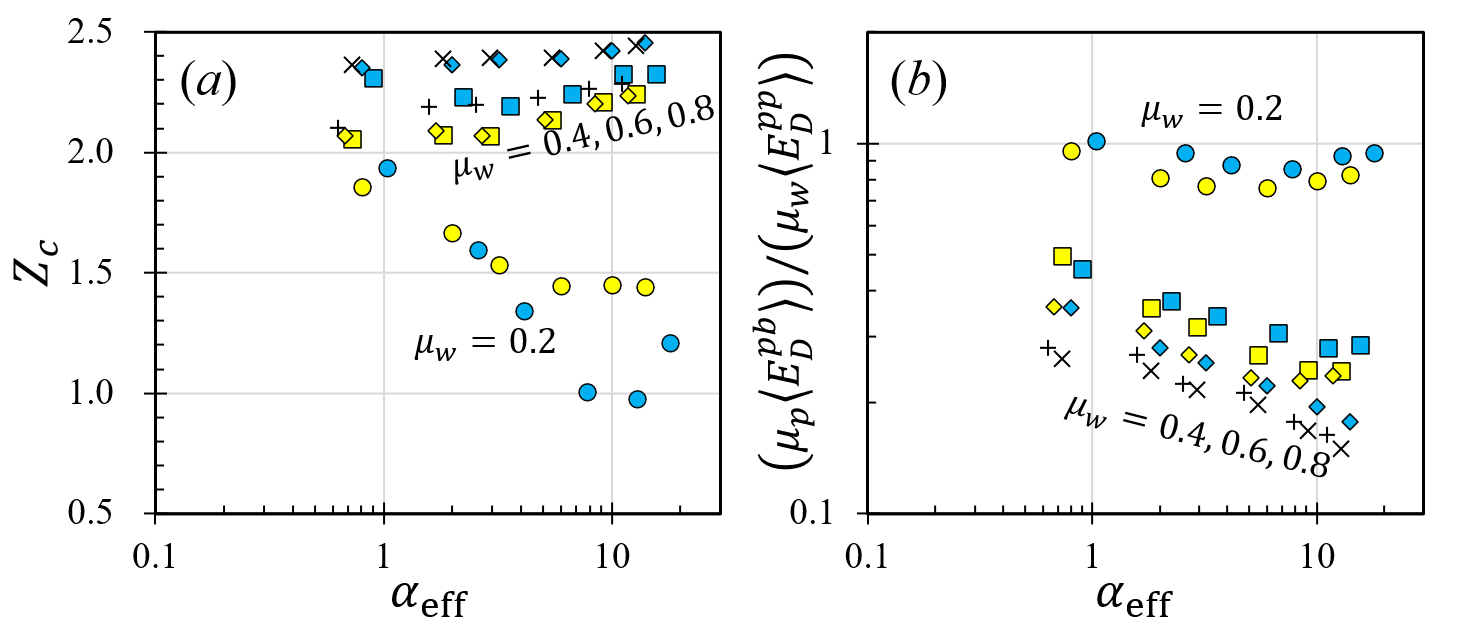}
  \caption{(a) The relationship between coordination number, $Z_c$, and $\alpha_{\textrm{eff}}$, where the coordination number, $Z_c$, stands for the average number of contacts a particle has in a granular assembly; (b) The relationship between $(\mu_p \langle E_{D}^{pb}\rangle)/(\mu_w \langle E_{D}^{pp}\rangle)$ and $\alpha_{\textrm{eff}}$, where $\langle E_{D}^{pb}\rangle$ is the time average of the energy dissipated by particle-boundary frictions, and $\langle E_{D}^{pp}\rangle$ stands for the time average of the energy dissipated by particle-particle frictions.}
  \label{figure7}
\end{figure}

To analyze the influence of inter-particle friction and particle/boundary friction separately, in Fig. \ref{figure7}(a) we plot the relationship between the final coordination number, $Z_c$, and $\alpha_{\textrm{eff}}$, where $Z_c$ stands for the average number of inter-particle contacts one particle has inside a granular assembly. When $\mu_w = 0.4, 0.6, 0.8$, $Z_c \gtrapprox 2.0$, which means that a cone-shape granular pile can be formed after the granular column collapse. However, when $\mu_w = 0.2$ and $\alpha_{\textrm{eff}}$ is large, $Z_c$ is decreased to between 1.0 and 1.5, which indicates that most particles are scattered on the plane with almost no inter-particle contacts. This fits our assumption that, when $\mu_w$ is small, the influence of particle/boundary interactions should be larger than other cases, hence, $\beta = 2.0$ might be no longer valid.

We then studied the ratio of the contribution from particle-boundary friction and that from inter-particle friction, and then connected this ratio of contributions with the energy dissipated by those two kinds of frictions, $(\mu_w\sigma_n)/(\beta\mu_p\sigma_n) \sim \langle E_{D}^{pb}\rangle/\langle E_{D}^{pp}\rangle$, where $\langle E_{D}^{pb}\rangle$ is the time average of the energy dissipated by particle-boundary frictions, and $\langle E_{D}^{pp}\rangle$ stands for the time average of the energy dissipated by particle-particle frictions, so $(1/\beta) \sim [(\mu_p \langle E_{D}^{pb}\rangle)/(\mu_w \langle E_{D}^{pp}\rangle)]$. We plotted $(\mu_p \langle E_{D}^{pb}\rangle)/(\mu_w \langle E_{D}^{pp}\rangle)$ against $\alpha_{\textrm{eff}}$ in Fig. \ref{figure7}(b). For most cases, this ratio remains less than 0.5, while for cases with $\mu_w = 0.2$, this ratio is much larger, which fits our assumption that the contribution of particle-boundary friction exceeds our expectation and the energy dissipation from particle-boundary friction becomes the major dissipation source when the particle-boundary frictional coefficient is sufficiently small. This also implies that, when particle/boundary friction is small, $\beta = 2$ could not hold. More detailed analyses on particle-scale contact forces between particles and the boundary should be particularly investigated to verify this hypothesis in the future.

We also hypothesize that the effective aspect ratio might reflect the ratio between the inertial effect and the frictional effect during the collapse of granular columns. According to Bagnold \cite{bagnold1954experiments,hill2011rheology}, in a dry granular flow, the inertial stresses scale with $f_1(\phi_s)\rho_p\dot{\gamma}^2d^2$, where $f_1(\phi_s)$ was a function of the solid fraction and he took the distance between adjacent particles in the flowing (or shearing) direction as the characteristic length. We interpret $\dot{\gamma}$ as a rate change of unit deformation, and $d$ as a length scale in the flowing direction. In our case, we simplify $\dot{\gamma}$ as $\sqrt{gH_i}/R_i$, which is the ratio between a characteristic velocity when particles fall over the length of $H_i$ and the length scale, $R_i$, over which the deformation occurs. Meanwhile, $d$ is replaced with $H_i$, giving the inertial stresses $\sigma_i \sim f_1(\phi_s)\rho_p(gH_i/R_i^2)H_i^2$. Additionally, the frictional stress can be calculated as $\sigma_f = \mu\sigma_n \sim \mu\phi_s\rho_pgH_i$, where $\phi_s$ is the solid fraction at the initial state. Thus the ratio between the two stresses can be obtained as
\begin{equation} \label{eq:stressRatio}
    \frac{\sigma_i}{\sigma_f} \sim \frac{f_1(\phi_s)\rho_p(gH_i/R_i^2)H_i^2}{\mu\phi_s\rho_pgH_i} = \frac{f_1(\phi_s)}{\mu\phi_s}\left(\frac{H_i}{R_i}\right)^2\ .
\end{equation}

If we regard the general frictional coefficient $\mu$ as a linear combination of the inter-particle frictional coefficient and the particle-boundary frictional coefficient, $\mu = \mu_w + \beta\mu_p$, with $\beta$ being a fitting parameter, and take the square root of this stress ratio, we can recover an effective aspect ratio with similar form as what we have obtained in Eq.\ref{eq:effect_alpha} and Eq.\ref{eq:g_mu},
\begin{equation} \label{eq:root_ratio}
    \alpha_{\textrm{eff}} = \sqrt{\frac{f_1(\phi_s)}{(\mu_w + \beta\mu_p)\phi_s}}\left(\frac{H_i}{R_i}\right)\ \ ,
\end{equation}
which suggests that the effective aspect ratio may have a deeper origin than what we have discovered in the dimensional analysis and could reflect a competition between different forms of stresses during the collapse of granular columns.

\subsection{Correlation analysis and regime transition}
\label{sec:regime}

\begin{figure*}
  \centering
  \includegraphics[width=0.75\textwidth]{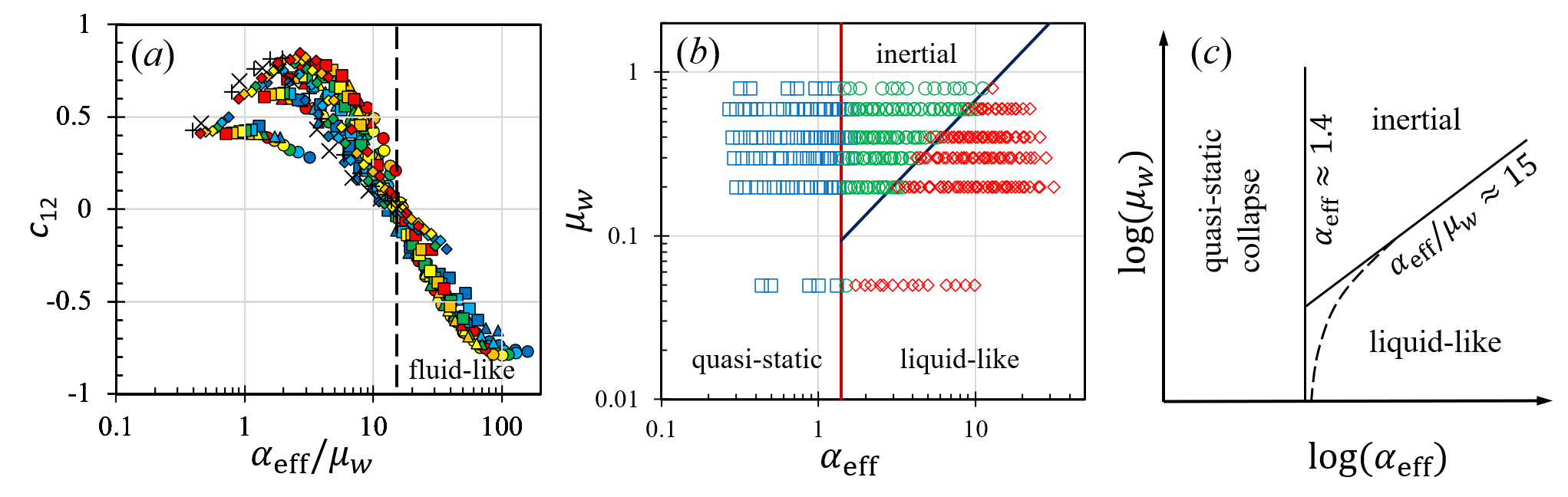}
  \caption{(a) plots the relationship between $c_{12}$ and $\alpha_{\textrm{eff}}/\mu_w$; (b) shows the transition among three collapse regimes. The solid lines in (b) indicate transition boundaries; (c) schematic diagram of the regime changes with respect to the effective aspect ratio $\alpha_{\textrm{eff}}$ and the boundary frictional coefficient $\mu_w$. The dash line in (c) shows the possible transition from an inertial regime to a liquid-like regime}
  \label{regime}
\end{figure*}

Figure\ \ref{effectRatio}(a) shows that the collapse of granular columns has two separate regimes based on the value of $\alpha_{\rm{eff}}$. However, according to the analysis of the deposition morphology, these granular flows should have three different regimes. To better investigate these regimes in connection to the deposition morphology, we analyze the relationship between the initial height, $h_i$, and the final radial position, $r_{\infty}$ (or the run-out distance, ${r}_{\infty} - {r}_i$) of each particle. Such a relationship can help us better understand how particles end up being in different positions. The correlation coefficient between $h_i$ and ${r}_{\infty} - {r}_i$ can then be obtained as
\begin{equation}
    \begin{split}
        c_{12} = \frac{\rm{cov}\it [h_i, ({r}_{\infty} - {r}_i)]}{\sigma_h \sigma_{r}}\ \ ,
    \end{split}
\end{equation}
where $c_{12}$ is the correlation coefficient between $r_{\infty}-r_i$ and $h_i$ of particles in a collapsed granular column, $\rm{cov}(\ )$ is the covariance function\cite{frangopol2008probability}, $\sigma_h$ is the standard deviation of all the initial height data in one system, and $\sigma_r$ is the standard deviation of all the run-out distances of particles in one system. When $c_{12}$ is positive, more particles initially at the top of a granular column end up at the foot of the deposited pile, whereas when $c_{12}$ is negative, more particles initially at the top cannot even reach the foot and more frequently end in the interior of the deposited granular packing, in which case, the collapse of granular columns behaves more like a fluid where top layer particles push the bottom layer particles away as they fall downward.

Figure\ \ref{effectRatio}(c) plots the relationship between $c_{12}$ and $\alpha_{\textrm{eff}}$. Both frictional coefficient and $\alpha_{\textrm{eff}}$ have great influence on how the $c_{12}$ evolves. When $\alpha_{\textrm{eff}}$ is small, $c_{12}$ is positive, and as we increase $\alpha_{\textrm{eff}}$, $c_{12}$ increases accordingly and eventually reaches a maximum, which corresponds to the turning point in the $\mathcal{R} - \alpha_{\rm{eff}}$ relationship in Fig. 5(a). The $\alpha_{\textrm{eff}}$ value that corresponds to the maxima of $c_{12}$ is a threshold which divides the quasi-static collapse and the inertial collapse. When $\alpha_{\textrm{eff}}$ is larger than this threshold, $c_{12}$ starts to decrease, and, at a certain point, it becomes negative, where the collapse of a granular column is in the liquid-like regime.

However, the point when systems change from an inertial regime to a liquid-like regime depends not only on $\alpha_{\rm{eff}}$ but also on other factors, as shown in Fig.\ \ref{effectRatio}(c), where the point of $c_{12}(\alpha_{\rm{eff}})$ crossing the $\alpha_{\textrm{eff}}$-axis is different for cases with different $\mu_w$. Interestingly, if we plot $c_{12}$ against $\alpha_{\rm{eff}}/\mu_w$ instead [Fig.\ \ref{regime}(a)], the place where $c_{12}$ passes zero collapses onto one point, $\alpha_{\rm{eff}}/\mu_w \approx 15$, which means that the transition from an inertial regime to a liquid-like regime depends on $\alpha_{\rm{eff}}/\mu_w$. Based on these analyses, a phase diagram of the collapse of granular columns can be obtained as Fig.\ \ref{regime}(b), where the collapse regime depends on both $\mu_w$ and $\alpha_{\rm{eff}}$. More specifically, the transition between quasi-static collapse and inertial collapse or liquid-like collapse depends only on the value of $\alpha_{\textrm{eff}}$. In contrast, the transition between the inertial regime and the liquid-like regime depends on the ratio between $\alpha_{\textrm{eff}}$ and $\mu_w$. It shows that the slope change in the $\mathcal{R}(\alpha_{\textrm{eff}})$ relationship (in log-log scale) corresponds to the regime change from quasi-static collapsing to the other two regimes. This change is solely governed by the effective aspect ratio. Physically, it indicates that the inertial force needs to be above a certain threshold to overcome the frictional force (linear combination of the inter-particle friction and particle-boundary friction). 

On the other hand, the change from an inertial collapse to a liquid-like collapse is mainly governed by $\alpha_{\textrm{eff}}/\mu_{p}$. When the granular column collapse is in a liquid-like regime, the curvature of the collapsed pile becomes non-monotonic and shows a more complex pattern. In this case, the initially top particles, instead of flowing to the front of the flow, end up flowing downward to push sub-layer particles to the front so that $c_{12}$ becomes negative. It should be noted that we also performed simulations with $\mu_w = 0.05$, and expected a direct change from quasi-static collapsing to liquid-like collapsing. However, there is still one case showing granular column collapse in the inertial regime [Fig. \ref{regime}(b)]. Thus, we suspect that when $\alpha_{\textrm{eff}}$ is approaching $\alpha_{\textrm{eff}} = 1.4$, the regime change between an inertial regime and a liquid-like regime should follow the dashed curve in Fig. \ref{regime}(c). It should also be noted that since our definition of $\alpha_{\textrm{eff}}$ includes the packing solid fraction, it can be potentially useful in describing the behavior of granular column collapse with different initial packing fractions. Details of the influence of packing fraction will be further investigated in future publications.

\section{Conclusions}
\label{sec:conclu}
In this paper, we analyzed the scaling of the run-out distance and deposition morphology of collapsed granular columns with respect to the aspect ratio based on numerical simulations. A dimensionless number $\alpha_{\textrm{eff}}$, which is derived from dimensional analysis and includes influences of the frictional coefficient (inter-particle and particle-boundary), the initial aspect ratio, and the initial solid fraction, is proposed. The $\alpha_{\textrm{eff}}$ can also be linked to the ratio between kinetic energy and collisionally dissipated energy. We showed that $\mathcal{R}-\alpha_{\textrm{eff}}$ relationships collapsed onto one curve for cases with different inter-particle and particle/boundary frictions. However, when $\mu_w = 0.2$ and $\alpha_{\textrm{eff}}$ is sufficiently large, the run-out distance is relatively larger than the prediction given by the mentioned collapsed curve. This is because when $\mu_w$ is small, particles tend to form a large area of one layer particles. During the formation of this one layer of particles, particle-boundary contacts dominate the sliding behavior of particles, while inter-particle contact becomes less frequent and less important. 

Based on the correlation between the initial position and final position of particles, we further classify the collapse of granular columns into three regimes (quasi-static, inertial, and liquid-like), which are controlled by both $\alpha_{\rm{eff}}$ and $\mu_w$. This study shows that the boundary friction can be crucial to the final deposition of a collapsed granular column, which brings insight to the behavior of landslides or debris flows in future studies. The introduction of the solid fraction in $\alpha_{\rm{eff}}$, although it plays no role in this study, could bring opportunities to understand better in coming studies how the initial packing of granular materials influences the collapsing dynamics of granular materials in both sub-aerial and sub-aqueous environment. Additionally, the results of this paper have led us to further explore the intriguing collapsing behavior when the cross-section of the granular column is no longer axisymmetric. It should be noted that granular materials have size effect. We find that the relative size of a granular column, $R_i/d$, could influence the run-out behavior (which was also observed by Warnett et al.\cite{warnett2014scalings} and Cabrera et al.\cite{Cabrera2019Granular}) and regime transitions. The different and exciting new results related to the influence of column size, column cross-section shape, as well as particle shapes will be presented in another publication.

%

\begin{acknowledgements}
The authors acknowledge the financial support from Westlake University and thank the Westlake University Supercomputer Center for computational resources and related assistance. T.M. would like to thank Prof. K. M. Hill from the University of Minnesota for helpful discussions. The authors specifically thank Prof. L. Staron for her helpful comments on an earlier version of this paper.
\end{acknowledgements}

%
\section*{Compliance with ethical standards}
\subsection*{Conflict of interest}
The authors declare that they have no conflict of interest.

\bibliographystyle{spmpsci}      
\bibliography{GranCollapseTM.bib}   

\end{document}